# Universality of galactic surface densities within one dark halo scale-length

Gianfranco Gentile[1,2], Benoit Famaey[2,3,4], HongSheng Zhao[5,6] & Paolo Salucci[7]

[1]Sterrenkundig Observatorium, Universiteit Gent, Krijgslaan 281, B-9000 Gent, Belgium. [2]Institut d'Astronomie et d'Astrophysique, Université Libre de Bruxelles, CP 226, Bvd du Triomphe, B-1050, Bruxelles, Belgium. [3]CNRS UMR 7550, Observatoire Astronomique, Université de Strasbourg, F-67000 Strasbourg, France. [4]AIfA, Universität Bonn, D-53121 Bonn, Germany. [5]SUPA, School of Physics and Astronomy, University of St Andrews, KY16 9SS, UK. [6]Leiden University, Sterrewacht and Instituut-Lorentz, Niels-Bohrweg 2, 2333 CA, Leiden, the Netherlands. [7]SISSA International School for Advanced Studies, via Beirut 4, I-34151, Trieste, Italy.

**It was recently discovered that the mean dark matter surface density within one dark halo scale-length (the radius within which the volume density profile of dark matter remains approximately flat) is constant across a wide range of galaxies[1]. This scaling relation holds for galaxies spanning a luminosity range of 14 magnitudes and the whole Hubble sequence[1–3]. Here we report that the luminous matter surface density is also constant within one scale-length of the dark halo. This means that the gravitational acceleration generated by the luminous component in galaxies is always the same at this radius. Although the total luminous-to-dark matter ratio is not constant, within one halo scale-length it is constant. Our finding can be interpreted as a close correlation between the enclosed surface densities of luminous and dark matter in galaxies[4].**

As a result of mass modelling using a Burkert dark matter halo[5], the density profile of which is defined by a constant central density $\rho_0$, and a scale-length (or core radius, at which the local dark matter volume density reaches a quarter of its central value) $r_0$, the product $\rho_0 r_0$ has been found to be the same for all galaxies[1]: $141^{+82}_{-52} \, M_\odot \text{pc}^{-2}$ , where $M_\odot \text{pc}^{-2}$ means solar masses per square parsec. The mean dark matter surface density within $r_0$, $\langle\Sigma\rangle_{0,\text{dark}}$, is then also constant when defined in terms of $M_{<r_0}$, the enclosed mass between $r = 0$ and $r = r_0$: $\langle \Sigma \rangle_{0,\,dark} = M_{<r_0} / \pi r_0^2 \approx 0.51 \rho_0 r_0 = 72^{+42}_{-27} \, M_\odot \mathrm{pc}^{-2}$. We note that the mean surface density as defined above is also proportional to the gravitational acceleration generated by the dark matter halo at $r_0$, which is also universal: $g_{dark}(r_0) = G\pi \langle \Sigma \rangle_{0,\,dark} = 3.2^{+1.8}_{-1.2} \, 10^{-9} \mathrm{cm} \, \mathrm{s}^{-2}$.

We sought to understand the intriguing universality[1–3] of the mean dark matter surface density within $r_0$ (and therefore of the gravitational acceleration due to dark matter at $r_0$). Hence, we searched for the dependency with galaxy magnitude of $g_{bary}(r_0)$, the gravitational acceleration due to baryons (that is, luminous matter) at $r_0$, and therefore also of $\langle \Sigma \rangle_{0,bary} = g_{bary}(r_0)/G\pi$, the baryonic mean surface density within $r_0$. In Fig. 1 we plot $g_{bary}(r_0)$ and $\langle \Sigma \rangle_{0,bary}$ as a function of galaxy magnitude. As in the case of dark matter, these values are universal: $g_{bary}(r_0) = 5.7^{+3.8}_{-2.8} \, 10^{-10} \mathrm{cm} \, \mathrm{s}^{-2}$. We note that, owing to deviation from spherical symmetry, the error bars on $\langle \Sigma \rangle_{0,bary}$ should be increased by at most 0.1 dex (decimal exponent, which is the unit of the logarithmic scale) in Fig. 1, but not the error bars on $g_{bary}(r_0)$.

This is therefore a new scaling relation relating baryonic parameters to dark halo parameters. Together with the universality of the dark matter surface density within $r_0$[1], it means that the baryonic-to-dark matter ratio is also universal within $r_0$. Our sample contains both small dark matter-dominated dwarf spheroidals and large late-type spirals, so this finding is not linked with a possible universality of the total baryonic-to-dark matter ratio within the sample. Neither does this result mean that the baryonic surface density of galaxies is constant, a misconception that used to be known as Freeman's law.

Because the central surface density of baryons actually varies by about four orders of magnitudes within the range of luminosities spanned in this study[1], our finding implies that the larger core radii $r_0$ of larger and more luminous galaxies compensate precisely for their larger baryonic surface densities to keep the same mean baryonic surface density within $r_0$. Whereas the universality of the dark matter surface density[1] within $r_0$ implies that the central dark matter density $\rho_0$ is precisely anti-correlated with the core radius, this new result means that, for a galaxy of a given size, the central baryonic surface density is actually correlated with the core radius. A large central luminous density thus implies a large core radius, and in turn a small central dark matter density. This precise balance must be the result of some unknown fine-tuned process in galaxy formation, because it is a priori difficult to envisage how such relations between dark and baryonic galaxy parameters can be achieved across galaxies that have experienced significantly different evolutionary histories, including numbers of mergers, baryon cooling or feedback from supernova-driven winds.

Another way of looking at this new relation is to say that the core radius of the dark matter halo is the radius beyond which the gravitational acceleration due to the baryons drops below a certain value, of the order of $6 \times 10^{-10}$ cm s$^{-2}$ (or $\log(g_{bary}) \approx -9.2$, where $g_{bary}$ is in centimetres per second squared). It is thus tempting to relate our finding to the mass discrepancy–acceleration relation[4,7,8], linking together the gravitational accelerations generated by the dark and luminous components at all radii in galaxies: other manifestations of this phenomenon are the universal rotation curve[9,10] (a parametrization of the rotation curves of spiral galaxies in which the rotation velocity depends only on radius and luminosity), or the radial Tully–Fisher[11] relation (a series of relations between luminosity and rotation velocity as a function of radius). Indeed, if the gravitational acceleration generated by the baryons is constant at $r_0$, it is a natural consequence of this mass discrepancy–acceleration relation that the baryonic-to-dark matter ratio should also be constant within this radius. Put another way, the mass discrepancy–acceleration relation means that the enclosed baryonic-to-dark matter ratio at any radius is uniquely correlated with the average surface density inside it[12].

To be more speculative, if we compare the effective dark matter profile predicted by the analytic form of the mass discrepancy–acceleration relation[7] with the enclosed mass profile of a Burkert halo[5] surrounding a point mass such that $g_{bary} = 5.7 \times 10^{-10}$ cm s$^{-2}$ at the core radius, we find that the two profiles are very similar in shape. Moreover, this effective dark matter profile typically predicts a universal maximum gravitational acceleration generated by the dark halo[13], of the order of $3 \times 10^{-9}$ cm s$^{-2}$. For any Burkert halo the maximum acceleration is generated precisely at $r_0$: from there, it then follows naturally that the gravity due to dark matter at $r_0$ should be universal and of that order of magnitude[1,13]. Then, from this universal and maximal acceleration due to dark matter at $r_0$, the mass discrepancy–acceleration relation (by definition) predicts that there is a universal gravity due to baryons at this radius, and thus a universal mean luminous surface density within it.

**Supplementary Information** is linked to the online version of the paper at www.nature.com/nature.

**Acknowledgements** G.G. was supported by the FWO-Vlaanderen (Belgium), and B.F. was supported in part by the Alexander von Humboldt foundation (Germany), the FNRS (Belgium), and the CNRS (France). We thank A. Jorissen and L. Hill for their comments on the manuscript.

**Author Contributions** G.G. and B.F. contributed to analysing and interpreting the data, making the figures, and writing the paper. HS.Z. and P.S. contributed to interpreting the data.

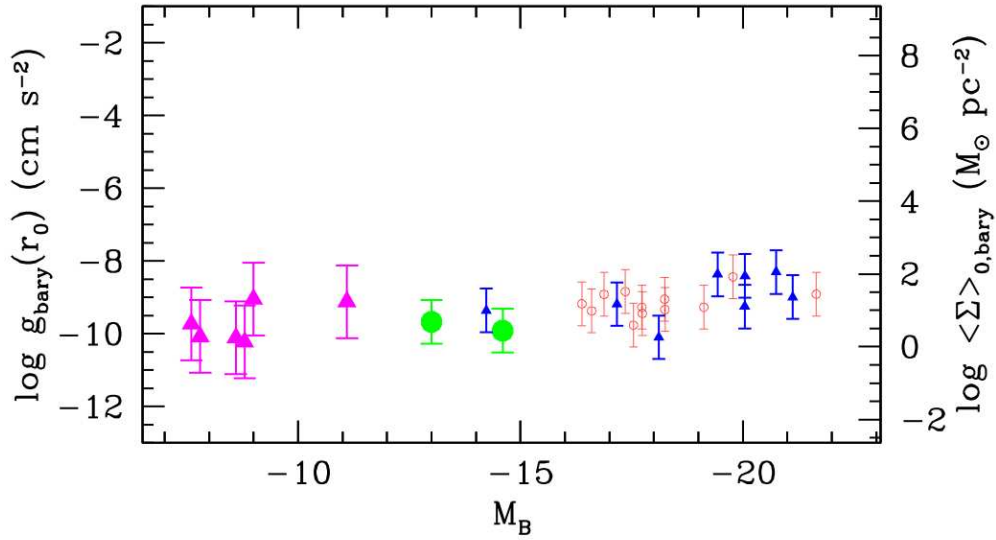

**Figure 1 Universality of the average surface density (and gravity) of baryons within the halo core radius.** $\langle\Sigma\rangle_{0,bary}$ and $g_{bary}(r_0)$ are plotted as a function of the B-band absolute magnitude of the galaxies. From the original sample[1], we used the dwarf spheroidals data, NGC 3741 (ref. 14), DDO 47 (ref. 15), and the two samples of spiral galaxies[3,16], which all together span the whole magnitude range probed in the original sample[1]. We excluded the galaxies that had uncertain values of $g_{bary}(r_0)$ owing to low-quality rotation curves, a low-quality fit, unrealistic stellar mass-to-light ratio ($\leq 0.1$), a core radius much larger than the last measured point of the rotation curve, or poor sampling of the data. The error bars on the calculated $g_{bary}(r_0)$ were based on a combination of the uncertainties on the stellar mass-to-light ratio and on the derived core radii (see Supplementary Information for details). The magenta triangles are the dwarf spheroidal galaxies[1], the left green point is NGC 3741 (ref. 14) and the right green point is DDO 47(ref. 15), and the empty red circles and filled blue triangles symbols are the Spano[3] and THINGS[16] spiral galaxies samples, respectively.

# Supplementary Notes

In this Supplementary Information, we give more details about the sample that was used to derive the main result of this Letter, about the individual galaxy parameters, and about the analytic form of the Mass Discrepancy-Acceleration relation. The first part describes how the sample was selected. The individual parameters of all the galaxies of the sample are listed in Supplementary Table 1. Then, Supplementary Figs. 1 and 2 show the scaling relations ensuing from our result and linking together various galaxy parameters. The second part of these supplementary notes explains how the error analysis was performed. Finally, the last part exhibits the link betweeen our result and the Mass Discrepancy-Acceleration relation (see also Supplementary Fig. 3).

## 1   The sample of galaxies

We started from the sample of galaxies[1] that was used for the discovery of the universality of the dark matter surface density within the core radius. However, for reasons outlined below, a number of galaxies of the original sample were not used. The present sample is composed by several subsamples: a sample of dwarf spheroidal galaxies[1,17] (dSphs), two dwarf irregular galaxies[14,15], and two samples of spiral galaxies of various Hubble types[3,16], leading to a total of 28 galaxies.

We eliminated a number of galaxies in the two samples of spirals, leaving only the ones which gave the most reliable value of the baryonic gravitational acceleration $g_{bary}$ at the core radius $r_0$. In the first sample[3], we eliminated eight galaxies because they were classified as having low-quality rotation curves (class 3 or 4) in the original paper[3] (UGC 4274, UGC 4325, UGC 5789, UGC 5842, UGC 7045, UGC 7699, UGC 7901, and UGC 10310); two galaxies an unrealistically low stellar M/L ratio of 0.1 (UGC 5279 and UGC 9219); one galaxy with a bad quality fit, as shown by the value of the reduced $\chi^2$ (UGC 9248); two galaxies with a rotation curve whose extent is less than three stellar exponential scale lengths (UGC 7876 and UGC 9866); three galaxies with a fit that did not converge unless by fixing some parameters (UGC 5175, UGC 6778, UGC 11914); six galaxies with a best-fit value of the core radius more than twice the last measured radius (UGC 2034, UGC 2455, UGC 2503, UGC 4555, UGC 5272, and UGC 6537), which reduces the first of the two samples of spirals to 12 galaxies (UGC 4256, UGC 4499, UGC 5721, UGC 7323, UGC 7524, UGC 8490, UGC 9179, UGC 9465, UGC 10075, UGC 11557, UGC 11707, and UGC 12060). The core radii given in the original paper[3], where a cored halo different from the Burkert halo was used, were converted using[1]: $r_0(\text{new}) = 1.26 \times r_0(\text{old})$.

From the second sample of spirals[1,16] we eliminated a galaxy with a low and uncertain inclination angle (NGC 6946), and two galaxies with an undefined value of the core radius (NGC 2976 and NGC 7331), leaving us with eight galaxies (DDO 154, NGC 925, NGC 2366, NGC 2403, IC 2574, NGC 3198, NGC 3621, and NGC 5055). Also in the case of this sample we converted the given core radii, this time following[1]: $r_0(\text{new}) = 1.74 \times r_0(\text{old})$. For this sample, since several fits with different values of the stellar M/L ratio were given, we derived $g_{bary}$ from the average stellar M/L ratios and core radii of these fits.

For the Ursa Minor dSph (one of the of six in the present sample), it has been shown[18] that the photometric fit from which the total luminosity was derived[19] lead to an underestimated luminosity: the corrected value is 2.7 times higher than the original one. Given that the luminosities of another two galaxies in the dSph sample (LeoII, and Sextans) were drawn from the same work[19], we conservatively multiplied their luminosity by 2.7. The remaining three dSphs (Carina, Draco, and LeoI) have deeper and more recent photometry[20–22], so we used these without any correction.

Finally, in the present analysis we took into account neither the weak lensing data (of a mixed sample of ellipticals and spirals) - because the data are sampled only every 45 kpc - nor the early-type spirals sample[23], because the uncertainties on the value of the core radius are larger than the value itself.

In Supplementary Table 1 we list the relevant observational properties and mass modelling parameters. Among the listed values is the dark halo core radius $r_0$. It is known to be degenerate with the stellar M/L ratio[24, 25] but its uncertainty is taken into account in the error budget of the acceleration due to baryons at $r_0$. Concerning the mass modelling parameters of the dSphs [17], isotropic Jeans modelling based on the whole velocity dispersion profile shows that the $r_0$ values are compatible with the scale length of the light distribution[26].

We then present two figures to illustrate the scaling relations between various galaxy parameters underpinning our result. Supplementary Fig. 1 illustrates how the universality of the mean dark matter surface density within $r_0$ links the central density $\rho_0$ with the core radius of the halo: it illustrates that $\rho_0$ is precisely anti-correlated with $r_0$. Supplementary Fig. 2 then illustrates how the larger core radii $r_0$ of larger and more luminous galaxies compensate for their larger baryonic surface densities. The scale radius $R_{\mathrm{p}}$ of the baryons was taken as the radius as which the baryonic contribution $V_{\mathrm{bary}}$ to the circular velocity reaches the peak[6]. We then measured the average baryonic surface density within this radius $(<\Sigma>(<R_{\mathrm{p}}))$. Among our sample, we eliminated in Supplementary Fig. 2 those galaxies where the determination of $R_{\mathrm{p}}$ was too uncertain (*i.e.*, where $V_{\mathrm{bary}}$ has a very wide plateau, spanning a large range of radii). For the dwarf spheroidal galaxies, $R_{\mathrm{p}}$ was derived from the equations describing the Plummer sphere, with which these galaxies were modelled. The correlation is not very tight, but the trend is clearly visible: galaxies with larger values of $<\Sigma>(<R_{\mathrm{p}})$ also have a larger value of the core radius. On the other hand, if we consider only galaxies having a relatively similar baryonic size, *e.g.* those with 5 kpc $< R_{\mathrm{p}} < 7$ kpc, the correlation is already a bit tighter.

Let us note that it is possible to find pairs of spiral galaxies with essentially identical asymptotic rotation velocities but very different baryonic surface densities. It is striking that the halo core radius then exactly compensates for the difference in baryonic density in order to keep the mean surface density universal inside $r_0$ without changing the outer rotation curve. Among the present sample, UGC 9179 and UGC 7323 both have an asymptotic velocity of $\sim 90$ km s$^{-1}$ but have a baryonic surface density $<\Sigma>(<R_{\mathrm{p}})$ equal to 8.7 $M_\odot$ pc$^{-2}$ and 38.9 $M_\odot$ pc$^{-2}$ respectively: their respective core radii are then 3.5 kpc and 9.7 kpc. As illustrated in Supplementary Fig. 1, they also have different central dark matter densities $\rho_0$, anti-correlated with the core radii (0.07 and 0.01 $M_\odot$ pc$^{-3}$ respectively). These different central dark matter densities take care of compensating for the different behaviour of the respective rotation curves at large radii linked with those different core radii in order not to affect the outer rotation curve while keeping a constant $\log(g_{\mathrm{bary}}) \approx -9.2$ at $r_0$.

## 2  Error analysis

The error bars on the value of the gravitational acceleration $g_{\mathrm{bary}}$ generated by the baryons at the core radius $r_0$ were generally determined from a combination of the uncertainties on the core radius and on the baryonic content within the core radius. These two uncertainties are not independent (the value of $g_{\mathrm{bary}}$ depends on the radius at which it is measured). However, taking full account of all the systematic uncertainties is practically extremely difficult, as it depends on difficultly quantifiable assumptions, like *e.g.* the shape of the velocity ellipsoid (in the case of the dSph) or the error bars of the observed rotation curves, which are known to be hard to assess in an objective manner.

In the case of the dSph, we estimated the uncertainty on $g_{\mathrm{bary}}$ to be as high as 1 dex, given (i) the typical uncertainty of a factor of 5 to 10 on the core radius[27,28] due to the degeneracy in the models caused by the unknown anisotropy of the velocity ellipsoid; and (ii) the uncertainty on the value of the stellar mass-to-light (M/L) ratio: we assumed a value of 2.65 for the stellar M/L ratio, intermediate between 1 and 4.3[29,30], the whole range 1–4.3 being representative of the uncertainty on the stellar M/L; (iii) the uncertainty on the luminosity itself, which can be as high as a factor of a few[18] (see also Section 1 of the Supplementary Information).

For the uncertainties on $g_{bary}$ for the disk galaxies, we added a typical uncertainty on the stellar M/L ratio (0.3 dex[31]), and a typical uncertainty on the core radius (which on average is of the same order of magnitude of the error on the stellar M/L ratio[32]). We did not derive the error on $g_{bary}$ from the quoted uncertainties in the original papers, as formal errors might underestimate the actual uncertainties on the fitted parameters. Indeed, errors of a few percent[32] are not representative of the actual uncertainties on the fitted parameters.

## 3  The Burkert halo and the Mass Discrepancy-Acceleration relation

Given the analytic form of the Mass Discrepancy-Acceleration relation[7,33,13] , we compared its predicted effective dark matter halo profile with a Burkert halo[5], which is known to give good fits to galaxy rotation curves. To have a unique dark matter profile prediction from the analytic form of the Mass Discrepancy-Acceleration relation, we considered the simplistic approximation where the baryonic distribution is approximated by a central point mass. One popular form of this analytic relation, giving at any radius the ratio of the enclosed mass of dark matter $M_{dark}$ to baryonic matter $M_{bary}$ as a function of the acceleration $g_{bary}$ generated by the baryons, is the following[7]:

$$\frac{M_{\text{dark}}}{M_{\text{bary}}} = \sqrt{\frac{1 + \sqrt{1 + (2a_0/g_{\text{bary}})^2}}{2}} - 1 \tag{1}$$

where $a_0 = 1.2 \times 10^{-8}$cm s$^{-2}$. We plotted this as a solid line in Fig. 3.

On the other hand, the Burkert halo[5] has the following density profile:

$$\rho_{\text{dark}}(r) = \frac{\rho_0 r_0^3}{(r + r_0)(r^2 + r_0^2)} \tag{2}$$

where $\rho_0$ is the central density, and $r_0$ the core radius. The enclosed mass $M_{\text{dark}}$ inside any radius $r$ is then given by:

$$M_{\text{dark}}(r) = 2\pi\rho_0 r_0^3 \left[\ln(1 + r/r_0) + \frac{1}{2}\ln(1 + r^2/r_0^2) - \arctan(r/r_0)\right]. \tag{3}$$

In the case of a central point mass $M_{\text{bary}}$, we fixed (to be consistent with our finding displayed in Fig. 1) the core radius $r_0$ as the radius at which the gravitational acceleration $g_{\text{bary}}$ generated by the baryons is $5.7 \times 10^{-10}$ cm s$^{-2}$:

$$r_0 = \sqrt{\frac{GM_{\text{bary}}}{5.7 \times 10^{-10}\,\text{cm}\,\text{s}^{-2}}}. \tag{4}$$

We then fixed the enclosed mass at the core radius $M_{\text{dark}}(r_0)$ [$\simeq 1.6\rho_0 r_0^3$] to be equal to the one derived from Eq. 1 for $g_{\text{bary}} = 5.7 \times 10^{-10}$ cm s$^{-2}$. This procedure fixed $\rho_0$, and thus the dark matter profile at *all* radii. The resulting enclosed mass profile from Eq. 3 is plotted as a dotted line in Supplementary Fig. 3. This mass profile is very similar to the one derived from Eq. 1 (solid line in Supplementary Fig. 3), which reveals the possible connection between our finding of a constant gravitational accleration at the core radius and the Mass Discrepancy-Acceleration relation. More extended baryon distributions will make the effective dark matter halo derived from the Mass Discrepancy-Acceleration relation less concentrated in the central parts, which will bring the two curves even closer.

More generally, irrespective of the baryonic distribution and of the precise analytic form of the Mass Discrepancy-Acceleration relation, Milgrom & Sanders[13] pointed out that this relation always predicts a maximum gravitational acceleration from the dark halo, generally of the order of $0.2a_0$ to $0.4a_0$ (and up to $a_0$ depending on the exact analytic form of the relation), *i.e.* $3^{+2}_{-1} \times 10^{-9}$ cm s$^{-2}$. Interestingly, the Burkert halo also predicts a maximum aceleration, which is reached precisely at $r_0$:

$$g_{\mathrm{dark_{max}}} = g_{\mathrm{dark}}(r_0) \simeq 1.6 G \rho_0 r_0 \qquad (5)$$

One would then expect the maximum halo acceleration predicted by the Mass Discrepancy-Acceleration relation to be close to the one of the Burkert halo (but, *e.g.* in Supplementary Fig. 3, not exactly at the same radius due to the slightly different shapes of the two profiles), meaning that the gravity due to baryons should be of the order of $3^{+2}_{-1} \times 10^{-9}$ cm s$^{-2}$ close to $r_0$, and that the product $\rho_0 r_0$ would then be compatible with the constant of Donato et al.[1], *i.e.* $\rho_0 r_0 = 132^{+88}_{-44}$ M$_\odot$ pc$^{-2}$. From there, a correlation as in Supplementary Fig. 1 of course follows naturally (see Fig. 4 of Milgrom & Sanders [13]).

# Supplementary References

| Name | $M_B$ | $r_0$ | $\log g_{\mathrm{bary}}(r_0)$ | $\log g_{\mathrm{dark}}(r_0)$ |
|---|---|---|---|---|
| | (mag) | (kpc) | (cm s$^{-2}$) | (cm s$^{-2}$) |
| Ursa Minor | -7.6 | 0.3 | -9.73 | -8.38 |
| Draco | -7.8 | 0.2 | -10.07 | -8.44 |
| Carina | -8.6 | 0.3 | -10.10 | -8.83 |
| Sextans | -8.8 | 0.7 | -10.23 | -9.29 |
| Leo II | -9.0 | 0.2 | -9.05 | -8.61 |
| Leo I | -11.1 | 0.3 | -9.11 | -8.50 |
| NGC 3741 | -13.1 | 3.0 | -9.67 | -8.79 |
| DDO 154 | -14.2 | 3.1 | -9.36 | -8.84 |
| DDO 47 | -14.6 | 6.9 | -9.92 | -8.65 |
| UGC 5721 | -16.4 | 1.3 | -9.18 | -8.04 |
| UGC 12060 | -16.6 | 5.8 | -9.37 | -8.57 |
| UGC 8490 | -16.9 | 2.4 | -8.92 | -8.39 |
| NGC 2366 | -17.2 | 2.2 | -9.19 | -8.19 |
| UGC 4499 | -17.4 | 2.8 | -8.84 | -8.67 |
| UGC 11707 | -17.5 | 3.0 | -9.76 | -8.35 |
| UGC 9179 | -17.7 | 3.5 | -9.27 | -8.36 |
| UGC 7524 | -17.7 | 4.3 | -9.44 | -8.74 |
| IC 2574 | -18.1 | 10.6 | -10.10 | -8.91 |

Supplementary Table 1: **List of galaxies and relevant parameters.** The columns are: 1) name; 2) B-band absolute magnitude; 3) dark halo core radius; 4) $g_{\mathrm{bary}}(r_0)$, the gravitational acceleration due to baryons at $r_0$; (for the error bars see Fig. 1 of the main part of the Letter and Section 2 of the Supplementary Notes); 5) $g_{\mathrm{dark}}(r_0)$, the gravitational acceleration due to dark matter at $r_0$. (see Donato et al. [1] for the error bars).

| Name | mag | $r_0$ | $\log g_{\mathrm{bary}}(r_0)$ | $\log g_{\mathrm{dark}}(r_0)$ |
|---|---|---|---|---|
| | | (kpc) | (cm s$^{-2}$) | (cm s$^{-2}$) |
| UGC 7323 | -18.2 | 9.7 | -9.05 | -8.68 |
| UGC 9465 | -18.3 | 16.4 | -9.34 | -8.54 |
| UGC 11557 | -19.1 | 6.3 | -9.27 | -8.70 |
| NGC 2403 | -19.4 | 4.3 | -8.37 | -8.22 |
| UGC 10075 | -19.8 | 3.5 | -8.43 | -7.94 |
| NGC 925 | -20.0 | 18.6 | -9.25 | -8.60 |
| NGC 3621 | -20.1 | 8.2 | -8.41 | -8.45 |
| NGC 3198 | -20.8 | 5.0 | -8.30 | -8.32 |
| NGC 5055 | -21.1 | 37.4 | -8.99 | -8.60 |
| UGC 4256 | -21.7 | 4.3 | -8.91 | -8.43 |

Supplementary Table 1: continued

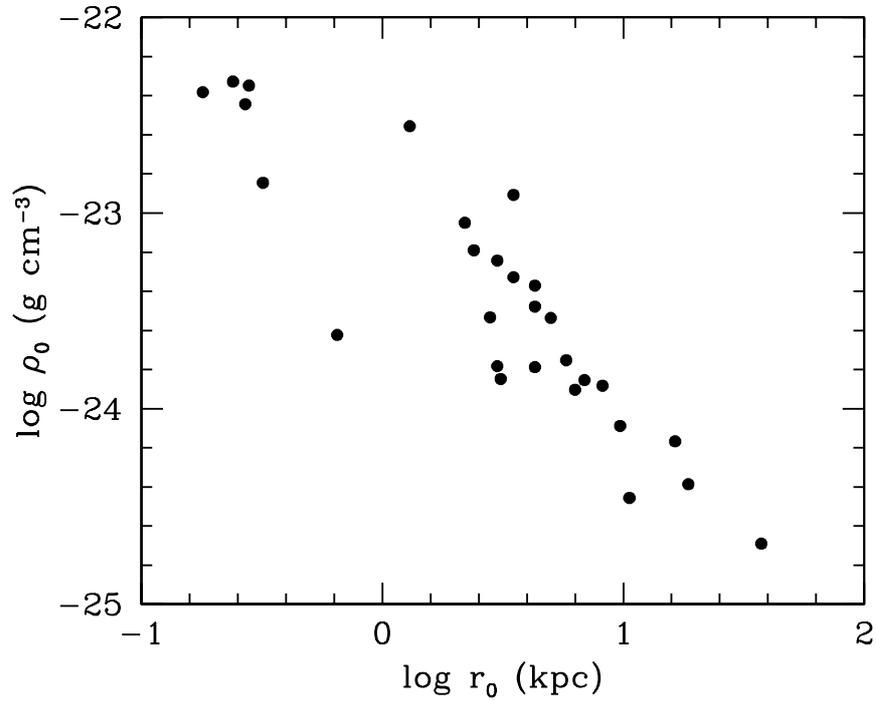

**Supplementary Figure 1: Central density of the dark halo ($\rho_0$) vs. its scale-length ($r_0$) for the galaxies of the present sample.** This clearly illustrates that the universality of the dark matter surface density within the core radius is due to this core radius being anti-correlated with the central dark matter density. Let us note that Kormendy & Freeman[2] found this relation (their Figs. 2, 3, and 4) with a larger sample of galaxies. Their Fig. 5 also shows that the scale-length times the central density is constant, which implies that each is inversely proportional to the other.

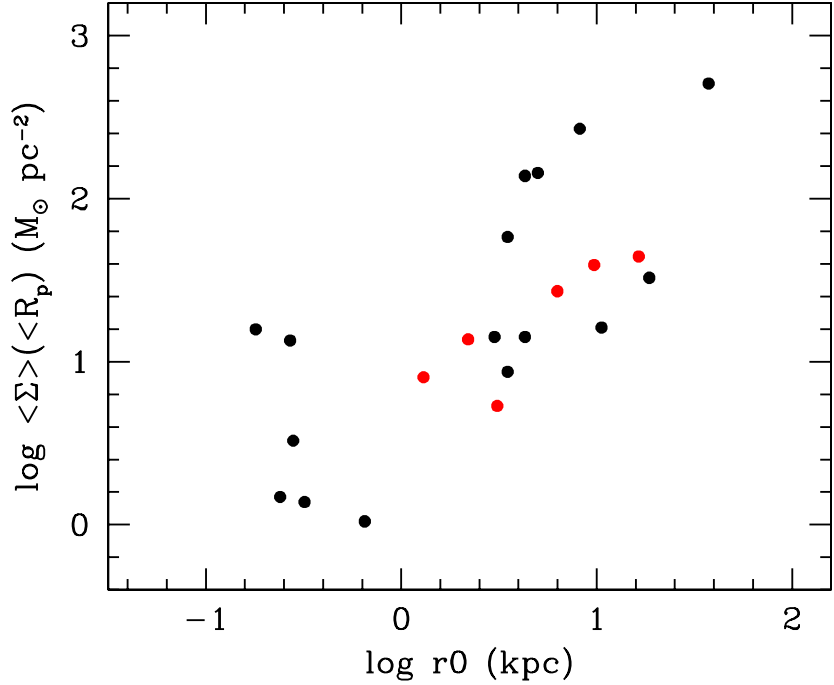

**Supplementary Figure** 2: **Average baryonic surface density vs. scale-length of the dark halo.** $< \Sigma > (< R_{\mathrm{p}})$ is the enclosed surface density within $R_{\mathrm{p}}$ (the radius at which the baryonic contribution $V_{\mathrm{bary}}$ to the rotation curve reaches a maximum), and $r_0$ is the scale-length of the Burkert halo. The galaxies with an uncertain $R_{\mathrm{p}}$ (*i.e.*, where $V_{\mathrm{bary}}$ has a very wide plateau) were removed. In red, we plot galaxies that have a similar $R_p$, 5 kpc $< R_{\mathrm{p}} < 7$ kpc: the correlation is clearly tighter.

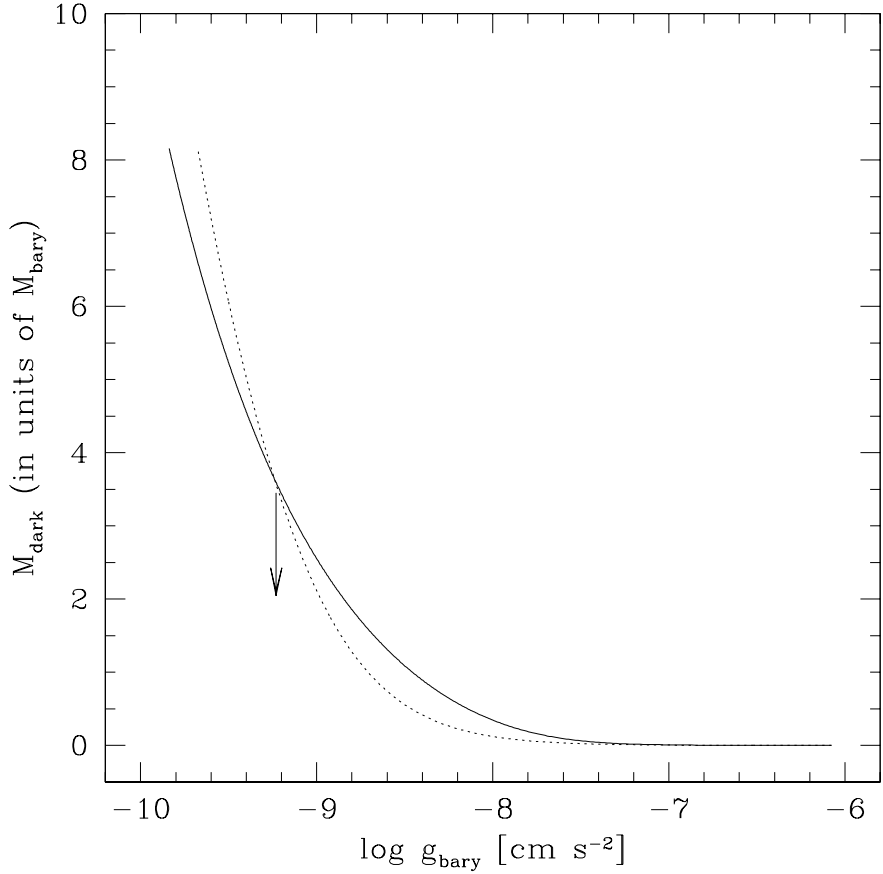

**Supplementary Figure** 3: **The Mass Discrepancy-Acceleration (MDA) relation and the Burkert halo.** The solid line is the analytical prediction[7] of the MDA relation (or MOND relation), giving the enclosed dark mass as a function of the gravitational acceleration generated by the baryons $g_{bary}$ in the special case of a central baryonic point mass $M_{bary}$. The dotted line is the enclosed mass of the Burkert halo, whose two parameters were chosen in such a way that $g_{bary} = 5.7 \times 10^{-10}$ cm s$^{-2}$ at the core radius $r_0$, and that at $r_0$ the enclosed dark matter masses are equal. The arrow indicates $g_{bary}$ at the core radius.